# Initializing the amplitude distribution of a quantum state

Dan Ventura* and Tony Martinez†
*Neural Networks and Machine Learning Laboratory‡,*
*Department of Computer Science, Brigham Young University, Provo, Utah 84602*
(Received 26 June 1998)

To date, quantum computational algorithms have operated on a superposition of all basis states of a quantum system. Typically, this is because it is assumed that some function $f$ is known and implementable as a unitary evolution. However, what if only some points of the function $f$ are known? It then becomes important to be able to encode only the knowledge that we have about $f$. This paper presents an algorithm that requires a polynomial number of elementary operations for initializing a quantum system to represent only the $m$ known points of a function $f$.

PACS numbers: 03.67.Lx, 03.67.-a, 89.80.+h, 03.65.Bz

quant-ph/9807054  18 Jul 1998

The study of computation as a physical process has produced fascinating results in the form of new approaches to problem solving and information processing that exhibit impressive speed-up over classical approaches for some problems [1--6]. All of these algorithms operate on a superposition of all basis states, usually assuming prior knowledge of some function $f$ that can be effected as a unitary evolution of the system. However, it is often the case in computational problem solving that a function $f$ is not known but rather that some example points of the function are all the knowledge available. Such situations are particularly common in the field of computational learning, which typically deals with learning a function from examples. If quantum computation is to become a broadly applicable paradigm, it must be possible for quantum algorithms to function given only a limited set of functional points. The first major step toward this goal is developing the ability to encode a set of examples of a function as the state of a quantum system. If this can be achieved, then the possibility of quantum computational learning algorithms presents itself. For example, fourier-based machine learning algorithms such as those by Kushilevitz and Mansour [7] and Jackson [8] are tempting candidates for generalization into the quantum realm. In fact, Bshouty and Jackson investigated a quantum computational approach to learning [9]. However, their algorithm lacks the capability proposed here (relying instead upon a *quantum example oracle* that somehow encodes the entire function $f$ even though the function is not known *a priori*) and therefore, though theoretically interesting, is not a plausible quantum algorithm. This paper presents an algorithm for encoding, in polynomial time, a set of examples of a function as the state of a quantum system.

Given a set $\mathcal{T}$ of $m$ examples of a function $f$, the goal is to produce

$$\left|\tilde{f}\right\rangle = \sum_{\bar{z}\in\mathcal{T}} f(\bar{z})|\bar{z}\rangle \qquad (1)$$

as the quantum state of $n$ qubits. It will be shown that the state $\left|\tilde{f}\right\rangle$ can be constructed using a polynomial number (in $n$ and $m$) of elementary operations on one, two, or three qubits. In what follows, the specific qubits to which an operator is to be applied are indicated as subscripts on that operator. For simplicity, consideration will first be restricted to the case of $f:\bar{z} \to s$ with $\bar{z} \in \{0,1\}^n$ and $s \in \{-1,1\}$. First, define the set of 2-qubit operators,

$$\hat{S}^{s,p} = \begin{bmatrix} 1 & 0 & 0 & 0 \\ 0 & 1 & 0 & 0 \\ 0 & 0 & \sqrt{\dfrac{p-1}{p}} & \dfrac{-s}{\sqrt{p}} \\ 0 & 0 & \dfrac{s}{\sqrt{p}} & \sqrt{\dfrac{p-1}{p}} \end{bmatrix}, \qquad (2)$$

where $s \in \{-1,1\}$ and $m \leq p \leq 1$. These operators form a set of conditional Hadamard-like transforms that will be used to incorporate the example set into a coherent quantum state. There will be a different $\hat{S}^{s,p}$ operator associated with each example in the set $\mathcal{T}$. Next, define

$$\hat{F} = \begin{bmatrix} 0 & 1 \\ 1 & 0 \end{bmatrix}, \qquad (3)$$

which flips the state of a qubit. Now, define the 2-qubit Control-NOT operator

$$\hat{F}^0 = \begin{bmatrix} \hat{F} & \hat{0} \\ \hat{0} & \hat{I}_2 \end{bmatrix}, \qquad (4)$$

where $\hat{0}$ and $\hat{I}_2$ are the 2×2 zero and identity matrices respectively, which conditionally flips the second qubit if the first qubit is in the $|0\rangle$ state. Similarly define $\hat{F}^1$ which conditionally flips the second qubit if the first qubit is in the $|1\rangle$ state [the matrix representing $\hat{F}^1$ is the same as Eq. (4) with $\hat{I}_2$ and $\hat{F}$ exchanged]. Finally, introduce four 3-qubit operators (really just four versions of the Fredkin gate [10]). These operators are used to identify specific states in a superposition, similar to Grover's identification of which state should be phase-rotated in his search algorithm [2]. The first of these,

$$\hat{A}^{00} = \begin{bmatrix} \hat{F} & \hat{0} \\ \hat{0} & \hat{I}_6 \end{bmatrix}, \qquad (5)$$

where the $\hat{0}$ are 6×2 and 2×6 zero matrices and $\hat{I}_6$ is the 6×6 identity matrix, conditionally flips the third bit if and only if the first two are in the state $|00\rangle$. Note that this is actually equivalent to $\hat{F}^{00}$, where $\hat{F}^{00}$ is the 3-qubit generalization of Eq. (4); however, this operator will be used with the third qubit always in the $|0\rangle$ state and thus can be thought of as performing a logical AND of the negation of the first two qubits, setting the third to $|1\rangle$ if and only if the first two are $|00\rangle$. The other 3-qubit operators, $\hat{A}^{01}$, $\hat{A}^{10}$, and $\hat{A}^{11}$, are variations of $\hat{A}^{00}$ in which $\hat{F}$ occurs in the other three possible locations along the main diagonal.

The algorithm for encoding $\mathcal{T}$ into a quantum system requires $n + (n - 1) + 2$ qubits, arranged in three quantum registers labeled $x$, $g$, and $c$. The quantum state of all three registers together is represented as $|x, g, c\rangle$, and the algorithm proceeds as in Fig. 1.

1. Generate $|\tilde{f}\rangle = |x_1...x_n, g_1...g_{n-1}, c_1c_2\rangle = |\bar{0}\rangle$
2. for $m \geq p \geq 1$
3.    for $1 \leq j \leq n$
4.      if $z_{pj} \neq z_{p+1\,j}$ (where $z_{m+1} = \{0\}^n$)
5.        $\hat{F}^0_{c_2 x_j} |\tilde{f}\rangle$
6.    $\hat{F}^0_{c_2 c_1} |\tilde{f}\rangle$
7.    $\hat{S}^{s_p, p}_{c_1 c_2} |\tilde{f}\rangle$
8.    $\hat{A}^{z_1 z_2}_{x_1 x_2 g_1} |\tilde{f}\rangle$
9.    for $3 \leq k \leq n$
10.      $\hat{A}^{z_k 1}_{x_k g_{k-2} g_{k-1}} |\tilde{f}\rangle$
11.    $\hat{F}^1_{g_{n-1} c_1} |\tilde{f}\rangle$
12.    for $n \geq k \geq 3$
13.      $\hat{A}^{z_k 1}_{x_k g_{k-2} g_{k-1}} |\tilde{f}\rangle$
14.    $\hat{A}^{z_1 z_2}_{x_1 x_2 g_1} |\tilde{f}\rangle$
15. $\hat{F}_{c_2} |\tilde{f}\rangle$

**FIG. 1. Initializing amplitude distribution in a quantum system**

The $x$ register will hold a superposition of the examples in the set $\mathcal{T}$ -- there are $n$ qubits in the register, and the value of $f$ for an example will be used as the coefficient for the state corresponding to that example. The $g$ and $c$ registers contain only ancillary qubits and are restored to the state $|\bar{0}\rangle$ by the end of the algorithm. A high-level intuitive description of the algorithm is as follows. The system is initially in the state $|\bar{0}\rangle$. The qubits in the $x$ register are conditionally flipped so that their states correspond to the first example. The $\hat{S}^{s,p}$ operator corresponding to that example then changes the basis of the system such that one of the states has a coefficient that matches the example's value for $f$. This same state then is changed to one that exists outside of the subspace affected by the $\hat{S}^{s,p}$ operators, in effect making it permanent, and the process is repeated for each example. When all the examples have been processed, the result is a coherent superposition of states corresponding to the sample points, where the amplitudes of the states all have the same magnitude but have different phases according to the values of their corresponding examples. This is somewhat reminiscent of the first part of Bernstein and Vazirani's fourier sampling mechanism [11]. However, they assume that the function $f$ is known and calculable in polynomial time, whereas here we do not know the function $f$, but only a small set of examples drawn from $f$.

In analyzing the complexity of the algorithm it is assumed that line 1 can be done trivially. The loop of line 2 is repeated $m$ times and consists of $n$ (lines 3-5) + 3 (6-8) + $n$-2 (9-10) + 1 (11) + $n$-2 (12-13) + 1 (14) operations, and line 15 requires one more operation. Thus, the entire algorithm requires $m(n + 3 + n\text{-}2 + 1 + n\text{-}2 + 1) + 1 = m(3n+1)+1$ operations and is $O(mn)$. This is optimal in the sense that just reading each instance once cannot be done in any fewer than $mn$ steps.

A concrete example of a simple 2-input function will help clarify the preceding discussion. Suppose that we are given the set $\mathcal{T} = \{f(01)=-1, f(10)=1, f(11)=-1\}$. The initial state is $|\bar{0}\rangle$, and the algorithm evolves the quantum state through the series of unitary operations described in Fig. 1. For convenience lines 3-6, 8-10, and 12-14 of the algorithm are agglomerated as the compound operators *FLIP*, *AND*, and *AND†* respectively and are treated in more detail in Eq. (6-8) and Figs. 2 and 3. First the qubit states of the $x$ register are flipped to match the first example, and the $c$ register is marked so the state will be affected by the $\hat{S}^{s,p}$ operator.

$$|00,0,00\rangle \xrightarrow{FLIP} |01,0,10\rangle$$

Next, the appropriate $\hat{S}^{s,p}$ operator (with $s$ equal to the output class of the instance being processed, here -1, and $p$ equal to the number of the instances, including the current one, yet to be processed, here 3) is applied, creating a new state in the superposition. This process will be referred to as *state generation* and corresponds to line 7 of Fig. 2.

$$\xrightarrow{\hat{S}^{-1,3}} -\tfrac{1}{\sqrt{3}}|01,0,11\rangle + \sqrt{\tfrac{2}{3}}|01,0,10\rangle$$

The two states just affected by the $\hat{S}^{s,p}$ operator are marked in their $g$ registers.

$$\xrightarrow{AND} -\tfrac{1}{\sqrt{3}}|01,1,11\rangle + \sqrt{\tfrac{2}{3}}|01,1,10\rangle$$

One of the marked states is made permanent by setting its $c$ register to $|01\rangle$ and this state now represents the first example. The $c$ register of the other state is returned to $|00\rangle$, and it is ready to generate a new state. This is performed by line 11 of the algorithm.

$$\xrightarrow{\hat{F}^1_{g_{n-1}c_1}} -\tfrac{1}{\sqrt{3}}|01,1,01\rangle + \sqrt{\tfrac{2}{3}}|01,1,00\rangle$$

Finally the work done in the $g$ register is undone and at this point one pass through the loop of line 2 of the algorithm has been performed.

$$\xrightarrow{AND^\dagger} -\tfrac{1}{\sqrt{3}}|01,0,01\rangle + \sqrt{\tfrac{2}{3}}|01,0,00\rangle$$

Now, the entire process is repeated for the second example. Again, the $x$ register of the appropriate state (that state whose $c_2$ qubit is in the $|0\rangle$ state) is selectively flipped to match the new example, and its $c$ register is again marked. This time the selective qubit state flipping occurs for those qubits that correspond to bits in which the first and second examples differ -- both in this case. Notice that the state representing the first example is not affected by this operation, nor any of those that follow.

$$\xrightarrow{FLIP} -\tfrac{1}{\sqrt{3}}|01,0,01\rangle + \sqrt{\tfrac{2}{3}}|10,0,10\rangle$$

Next, another $\hat{S}^{s,p}$ operator is applied, this time with $s = 1$ and $p = 2$.

$$\xrightarrow{\hat{S}^{1,2}} -\tfrac{1}{\sqrt{3}}|01,0,01\rangle + \tfrac{1}{\sqrt{2}}\sqrt{\tfrac{2}{3}}|10,0,11\rangle + \sqrt{\tfrac{1}{2}}\sqrt{\tfrac{2}{3}}|10,0,10\rangle$$

The two states just affected by the $\hat{S}^{s,p}$ operator are marked in their $g$ registers.

$$\xrightarrow{AND} -\tfrac{1}{\sqrt{3}}|01,0,01\rangle + \tfrac{1}{\sqrt{3}}|10,1,11\rangle + \sqrt{\tfrac{1}{3}}|10,1,10\rangle$$

The $c$ register is used to save another state (this one representing the second example).

$$\xrightarrow{\hat{F}^1_{g_{n-1}c_1}} -\tfrac{1}{\sqrt{3}}|01,0,01\rangle + \tfrac{1}{\sqrt{3}}|10,1,01\rangle + \sqrt{\tfrac{1}{3}}|10,1,00\rangle$$

A final bit of house keeping resets the $g$ register, preparing for the process to be repeated again.

$$\xrightarrow{AND^\dagger} -\tfrac{1}{\sqrt{3}}|01,0,01\rangle + \tfrac{1}{\sqrt{3}}|10,0,01\rangle + \sqrt{\tfrac{1}{3}}|10,0,00\rangle$$

Finally, the third example is considered. The *x* register of the generator state is again selectively flipped and its *c* register is marked. This time, only those qubits corresponding to bits that differ in the second and third examples are flipped, in this case just qubit $x_2$.

$$\xrightarrow{FLIP} -\tfrac{1}{\sqrt{3}}|01,0,01\rangle + \tfrac{1}{\sqrt{3}}|10,0,01\rangle + \sqrt{\tfrac{1}{3}}|11,0,10\rangle$$

Again a new state is generated to represent this third example. Notice that the generator state is now left with a magnitude of 0, because there are no more examples to process.

$$\xrightarrow{\hat{S}^{-1,1}} -\tfrac{1}{\sqrt{3}}|01,0,01\rangle + \tfrac{1}{\sqrt{3}}|10,0,01\rangle - \tfrac{1}{\sqrt{1}}\sqrt{\tfrac{1}{3}}|11,0,11\rangle + \sqrt{\tfrac{0}{1}}\sqrt{\tfrac{1}{3}}|11,0,10\rangle$$

Once again the house keeping steps are performed.

$$\xrightarrow{AND} -\tfrac{1}{\sqrt{3}}|01,0,01\rangle + \tfrac{1}{\sqrt{3}}|10,0,01\rangle - \tfrac{1}{\sqrt{3}}|11,1,11\rangle$$

$$\xrightarrow{\hat{F}^1_{g_m c_1}} -\tfrac{1}{\sqrt{3}}|01,0,01\rangle + \tfrac{1}{\sqrt{3}}|10,0,01\rangle - \tfrac{1}{\sqrt{3}}|11,1,01\rangle$$

$$\xrightarrow{AND^\dagger} -\tfrac{1}{\sqrt{3}}|01,0,01\rangle + \tfrac{1}{\sqrt{3}}|10,0,01\rangle - \tfrac{1}{\sqrt{3}}|11,0,01\rangle$$

Finally, line 15 performs one last operation, restoring all the ancillary qubits to their initial state.

$$\xrightarrow{F_{c_2}} -\tfrac{1}{\sqrt{3}}|01,0,00\rangle + \tfrac{1}{\sqrt{3}}|10,0,00\rangle - \tfrac{1}{\sqrt{3}}|11,0,00\rangle$$

At this point, the *g* and *c* registers are not entangled with the *x* register, and therefore the system can be simplified as

$$-\tfrac{1}{\sqrt{3}}|01\rangle + \tfrac{1}{\sqrt{3}}|10\rangle - \tfrac{1}{\sqrt{3}}|11\rangle,$$

and it may be seen that the partial function defined by the set $\mathcal{T}$ is now represented as a quantum superposition in the *x* register.

Here the complex operators *FLIP*, *AND*, and $AND^\dagger$ are specified in more detail. In operator notation,

$$FLIP = \hat{F}^0_{c_2 c_1} \hat{F}^0_{c_2 x_j} \left(n \geq j \geq 1,\ z_{pj} \neq z_{p+1,j}\right). \tag{6}$$

Alternatively, *FLIP* may be represented as a quantum network as in [12]. Figure 2 shows one such network for a 3-input function. In the figure, a '?' indicates that the $\hat{F}^0$ operator is applied only if the value of the bit in question differs from the value of that bit for the previous example.

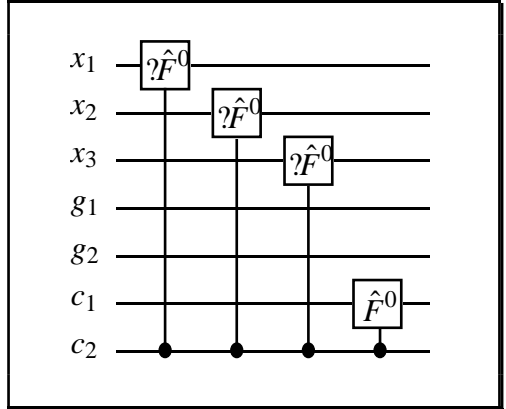

FIG. 2. Quantum network for implementing *FLIP*

Also, in operator notation

$$AND = \hat{A}^{z_n 1}_{x_n g_{n-2} g_{n-1}} \cdots \hat{A}^{z_3 1}_{x_3 g_1 g_2} \hat{A}^{z_1 z_2}_{x_1 x_2 g_1} \tag{7}$$

and

$$AND^\dagger = \hat{A}^{z_1 z_2}_{x_1 x_2 g_1} \hat{A}^{z_3 1}_{x_3 g_1 g_2} \cdots \hat{A}^{z_n 1}_{x_n g_{n-2} g_{n-1}}. \tag{8}$$

*AND* and *AND*$^\dagger$ also may be represented as quantum networks, and corresponding example diagrams appear in Figs. 3(a) and 3(b) respectively. Note that the only difference between the two is the order in which the qubits are operated upon since all the $\hat{A}$ operators are their own adjoints.

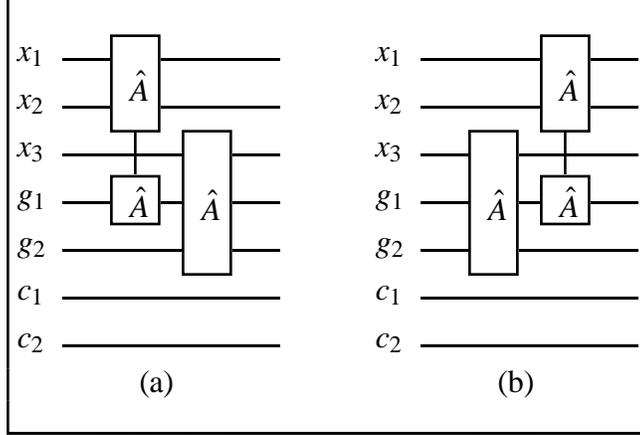

**FIG. 3. Quantum networks for implementing *AND* and *AND*$^\dagger$**

The quantum network representation of the entire algorithm requires the addition of gates for performing the $\hat{S}$ and $\hat{F}$ operations, and the complete network is shown in Fig. 4. The network shown will be repeated several times -- once for each instance in the training set. Each such repetition will make use of different $\hat{A}$ operators and flip different qubits using the $\hat{F}$ gates, according to the particular example being processed.

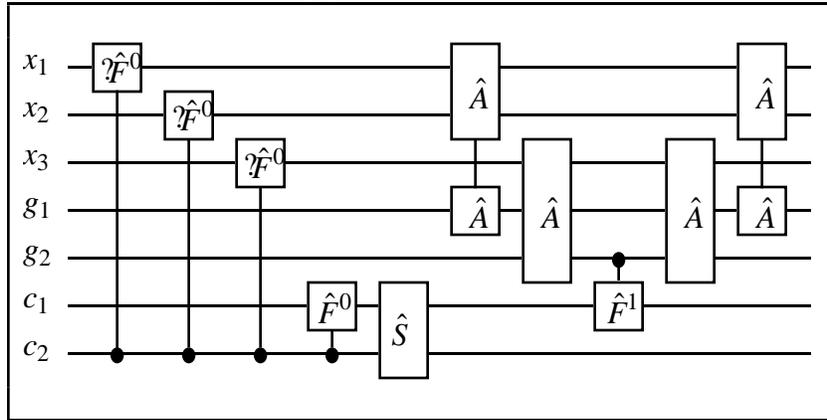

**FIG. 4. Quantum network for encoding a single function example**

The trick that makes the algorithm work is the fact that whenever an $\hat{S}^{s,p}$ operator is applied, there are no states with the $c$ register in the $|11\rangle$ state. This is crucial to the construction of $|\tilde{f}\rangle$ because it allows the operations to be unitary without generating "extraneous" states. Another key is the ability to identify a specific state in the superposition (the one just generated) using the *AND* and $AND^\dagger$ operators; this provides the ability to mark that particular state and only that state so as not to operate on it again. And this is done reversibly so that no disruptive entanglement results.

The algorithm of Fig. 1 can handle only binary functions. However, generalization to functions with $N$ input/output values is straightforward. The only necessary changes are a generalization of the $x$ register (and only the $x$ register) to include systems with more than two states, generalization of some of the operators and the representation of more than two output values as points on the unit circle in the complex plane (representing a binary output with the values -1 and 1 is just a special case of this). The operator generalizations are as follows.

$$\hat{S}^{s,p} = \begin{bmatrix} 1 & 0 & 0 & 0 \\ 0 & 1 & 0 & 0 \\ 0 & 0 & \sqrt{\frac{p-1}{p}} & \frac{-e^{(s2\pi i/N)}}{\sqrt{p}} \\ 0 & 0 & \frac{e^{(s2\pi i/N)}}{\sqrt{p}} & \sqrt{\frac{p-1}{p}} \end{bmatrix}, \quad (9)$$

where now $s \in \{0, ..., N-1\}$ and $m \leq p \leq 1$. The unit circle in the complex plain is partitioned into $N$ pie pieces, and the points on the pie piece borders correspond to the function values. By convention, one of the values is always on the positive real line.

$$^N\hat{F}^0 = \begin{bmatrix} \hat{C} & \hat{0} \\ \hat{0} & \hat{I}_N \end{bmatrix}, \quad (10)$$

where $\hat{0}$ and $\hat{I}_N$ are the $N \times N$ zero and identity matrices respectively, and $\hat{C}$ is the $N \times N$ circulant matrix obtained by shifting all rows of $\hat{I}_N$ down one, with the $N$th row becoming the first. There is now a set of $\hat{F}^0$ operators, one for 2-state qubits, one for 3-state qubits, etc., and the $\hat{F}^0$ operator defined in Eq. (4) is the special case for $N=2$. These operators act on a pair of qubits, the first of which is in the $c$ register and thus is still just a 2-state system, and the second of which is in the $x$ register and thus now has $N$ states, as $^N\hat{F}^0|0q\rangle = |0(q+1) \mod N\rangle$ and $^N\hat{F}^0|1q\rangle = |1q\rangle$, conditionally rotating the second qubit (labeled $q$) to the next highest value (but rotating the maximum value back to 0) if the first qubit is in the $|0\rangle$ state. Note that $\hat{F}^1$ does not need to be generalized since it only acts on qubits in the $g$ and $c$ registers, which are still only 2-state qubits.

$$^N\hat{A}^{00} = \begin{bmatrix} \hat{F} & \hat{0} \\ \hat{0} & \hat{I}_a \end{bmatrix}, \quad (11)$$

where $a = 2N^2-2$, the $\hat{0}$ are $2 \times a$ and $a \times 2$ zero matrices, $\hat{I}_a$ is the $a \times a$ identity matrix, $\hat{F}$ is as defined in Eq. (3) and the entire matrix is $2N^2 \times 2N^2$. Again there is now a set of $\hat{A}^{00}$ operators and the $\hat{A}^{00}$ operator defined in Eq. (5) is the special case for $N=2$. These operators act on three qubits, as $^N\hat{A}^{00}|00r\rangle = |00(r+1) \mod 2\rangle$ and $^N\hat{A}^{00}|pqr\rangle = |pqr\rangle$ for $p,q \neq 0$, conditionally flipping the third qubit (labeled $r$), which will still be a 2-state qubit as it is in the $g$ register, if the first two qubits are in the $|00\rangle$ state. Note that whereas the operator $\hat{A}^{00}$ defined in Eq. (5) was really equivalent to a 3-qubit version of Eq. (4), this is *not* the case with these generalized operators. In other words, Eq. (11) is not equivalent to a 3-qubit generalization of Eq. (10). This is because $^N\hat{F}^0$ now affects qubits (in the $x$ register) with more than 2 states, therefore requiring a rotation through the various states rather than a simple flip between 2 possible states; whereas $^N\hat{A}^{00}$ still affects qubits with only 2 states and thus simply toggles between states. Since this operator will be used with the third bit always in the $|0\rangle$ state, it can still be thought of as performing a sort of logical AND (actually a generalized AND where the inputs can be greater than binary) on the values of the first two qubits, setting the third to $|1\rangle$ if and only if the first two are $|00\rangle$. The other $\hat{A}$ operators must also be generalized, a set for each possible pair of values that any of the qubits in the *x* and *g* registers can assume. In other words we need all the sets of operators $^N\hat{A}^{ij}$ for $0 \leq i,j \leq N\text{-}1$. However, these sets of operators are simple variations of the $^N\hat{A}^{00}$ set given in Eq. (11) with the sub-matrix $\hat{F}$ occurring in different locations along the diagonal according to the values for *i* and *j*.

The algorithm for the more general case is the same as in Fig. 1, except for two changes. First, the operators must be replaced with their generalizations defined in Eq. (9-11), and second, the IF statement of line 4 must be changed to a WHILE statement. This is again because instead of simply flipping a qubit's state between two possibilities as the binary operator $\hat{F}^0$ defined in Eq. (4) does, the generalized operator $^N\hat{F}^0$ defined in Eq. (10) rotates a qubit's state through *N* possibilities in a specific order one at a time. Thus, the compound *FLIP* operation defined by lines 3-6 of the algorithm now may require up to $(N\text{-}1)n$ operations instead of the $O(n)$ required for the simpler binary case. This brings the total number of operations required for the algorithm up to $(N+1)mn + m + 1$, which is still $O(mnN)$. This reduces to $O(mn)$ assuming $N << m,n$.

In summary, this paper presents a polynomial-time algorithm for initializing a quantum system to represent *m* known points of a function *f*. The result is a quantum superposition with *m* non-zero coefficients -- the creation of which is a nontrivial task compared to creating a superposition of all basis states. It may be appropriate to mention here that very recently work as been done to analyze Grover's algorithm [2] for the case of arbitrary initial amplitude distributions [13]. Further, the paper suggests a new field to which quantum computation may be applied to advantage -- that of computational learning. In fact, it is the authors' opinion that this application of quantum computation will, in general, demonstrate much greater returns than its application to more traditional computational tasks (though Shor's algorithm is an obvious exception). We make this conjecture because results in both quantum computation and computational learning are by nature probabilistic and inexact, whereas most traditional computational tasks require precise and deterministic outcomes.

*dan@axon.cs.byu.edu
†martinez@cs.byu.edu
‡http://axon.cs.byu.edu

[1] P. Shor, SIAM J. Comput. **26**, 1484 (1997).
[2] L. Grover, *Proceedings of the 28th ACM Symposium on the Theory of Computing*, *Philadelphia*, *1996*, edited by Gary L. Miller (ACM Press, 1996).
[3] D. Simon, SIAM J. Comput. **26**, 1474 (1997).
[4] D. Deutsch and R. Jozsa, Proc. Roy. Soc. London Ser. A **439**, 553 (1992).
[5] T. Hogg, Journal of Artificial Intelligence Research **4**, 91 (1996).
[6] B. M. Terhal and J. A. Smolin, Phys. Rev. A (to be published).
[7] E. Kushilevitz and Y. Mansour, SIAM J. of Comput. **22**, 1331 (1993).
[8] J. Jackson, Journal of Computer and System Sciences **55**, 414 (1997).
[9] N. H. Bshouty and J. Jackson, in *Proceedings of the 8th Annual Conference on Computational Learning Theory*, *Santa Cruz*, *1995*, edited by Wolfgang Maass (ACM Press, 1995).
[10] E. Fredkin and T. Toffoli, Internat. J. Theoret. Phys. **21**, 219 (1982).

[11]  E. Bernstein and U. Vazirani, SIAM J. Comput. **26**, 1411 (1997).
[12]  D. Deutsch, Proc. Roy. Soc. London Ser. A **425**, 73 (1989).
[13]  D. Biron *et al.*, in *Proceedings of the 1st NASA International Conference on Quantum Computing and Quantum Communications*, *Palm Springs*, *1998*, edited by Charles Bennett, (to be published).